# Majority of water hides deep in the interiors of exoplanets

Haiyang Luo[1*†], Caroline Dorn[2*†] and Jie Deng[1*]

[1]Department of Geosciences, Princeton University, Princeton, 08544 New Jersey, USA.
[2]Institute for Particle Physics and Astrophysics, ETH Zürich, Otto-Stern-Weg 5, 8093 Zürich, Switzerland.
*Corresponding authors. Email: haiyang.luo@princeton.edu, dornc@ethz.ch, or jie.deng@princeton.edu
[†]These two authors contribute equally to this work.

**Abstract:** Water is an important component of exoplanets, with its distribution, i.e., whether at the surface or deep inside, fundamentally influencing the planetary properties. The distribution of water in most exoplanets is determined by yet-unknown partitioning coefficients at extreme conditions. Our new *ab initio* molecular dynamics simulations reveal that water strongly partitions into iron over silicate at high pressures and thus would preferentially stay in a planet's core. Furthermore, we model planet interiors by considering the effect of water on density, melting temperature, and water partitioning. The results shatter the notion of water worlds as imagined before: the majority of the bulk water budget (even more than 95%) can be stored deep within the core and the mantle, and not at the surface. For planets more massive than ~6 $M_\oplus$ and Earth-size planets (of lower mass and small water budgets), the majority of water resides deep in the cores of planets. Whether water is assumed to be at the surface or at depth can affect the radius by up to 25% for a given mass. This has drastic consequences for the inferred water distribution in exoplanets from mass-radius data.

**Main text:** The majority of detected super-Earths and sub-Neptunes are expected to have been hosting global magma oceans since formation, due to their high initial temperatures and limited cooling resulting from intense irradiation from host stars. Magma oceans serve as crucial reservoirs for storing volatiles, particularly water, which exhibits higher solubility compared to other gas species[1]. Hence, water can be present deep within planetary interiors[2]. Additionally, water-rich worlds are predicted by planet formation theories[3], further supporting the abundance of water. While the dissolution of volatiles has been considered in recent years in theoretical studies on interior modelling[4-6], the majority of present-day interior models used for interpreting exoplanet mass-radius data commonly assume that all water exists solely at the surface of a rocky interior, neglecting the presence of water deep within planetary interiors. Efforts to reach high accuracy in observed mass-radius data must be matched with precise theoretical interior models. This involves developing a comprehensive and realistic description of how different planet building materials, such as iron, silicate, water, and gas, interact.

Here we focus on water partitioning between an iron core, a silicate mantle, and a surface reservoir. The water partitioning behavior between silicate melt and vapor has been well-established[7]. Notably, a previous study suggests that hidden water in magma oceans may result in a deviation in planet radius of up to 16% given a fixed planet mass and bulk composition[8]. No studies have yet explored the water partitioning between iron and silicate melt under the extreme pressure and temperature conditions relevant to super-Earths and sub-Neptunes. Water in the metallic core is especially important because these waters will be locked away for the lifetime of the planet, while water in the mantle can be outgassed over time. Taking Earth as an essential reference point for studying rocky exoplanets, recent studies indicate a strong influence of Earth's core on the distribution of water and suggest that 37-73 Earth ocean masses of water are stored in the core as water has large iron/silicate partition coefficients ($D_{\mathrm{H_2O}}^{\mathrm{iron/silicate}}$= water mass fraction in iron/water mass fraction in silicate ≥ 29) at core formation conditions (up to ~135 GPa & 4200 K)[9-11]. Super-Earths and sub-Neptunes with larger masses and surface gravities experience higher pressures (up to ~1400 GPa) and temperatures (up to ~14000 K) during their core formation. Without the data obtained at the relevant *P/T* conditions, it is difficult to estimate the metal-silicate partitioning behavior of water at the conditions of super-Earths and sub-Neptunes, given the ~10 times larger pressures and the induced drastic structural changes of metal and silicate. In fact, extrapolated values[9,11] of $D_{\mathrm{H_2O}}$ at these conditions remain inconsistent with each other and span a wide range from 0 to $10^{11}$ (Supplementary Fig. 1), rendering the water distribution in super-Earths and sub-Neptunes largely unconstrained.

In this study, we calculate partition coefficients of $H_2O$ between iron and silicate melt ($D_{H_2O}^{Fe/MgSiO_3}$) using *ab initio* molecular dynamics simulations coupled with the thermodynamic integration (TI) technique (Methods). This method has been successfully applied to predict the metal-silicate partitioning of many elements, including hydrogen[9], boron[12], and noble gases[13,14]. We initiate our calculations at three hypothetical conditions for super-Earths: 250 GPa/6500 K, 500 GPa/9000 K, and 1000 GPa/13000 K (Extended Data Fig. 1 & Extended Data Table 1). We then investigate the temperature dependence of $H_2O$ partitioning based on the Gibbs-Helmholtz equation by calculating enthalpies at various temperatures at fixed pressures (Extended Data Fig. 2 and 3 & Extended Data Table 2). In addition to the TI method, we also perform direct two-phase molecular dynamic simulations of metal-silicate coexistence to gain further insight into the lithophile or siderophile nature of water (Extended Data Fig. 4).

Our results based on the TI method reveal that under the extreme conditions of super-Earth and sub-Neptune, $H_2O$ remains siderophile and exhibits comparable partition coefficients $D_{H_2O}^{Fe/MgSiO_3}$ at 250 GPa/6500 K, 500 GPa/9000 K, and 1000 GPa/13000 K within error (Fig. 1a). $D_{H_2O}^{Fe/MgSiO_3}$ decreases with concentration of $H_2O$ in iron melt ($X_{H_2O}^{Fe}$) and plateaus when $X_{H_2O}^{Fe}$ exceeds around 7 wt%. This is because more $H_2O$ tends to enter iron melt to maximize the mixing entropy when the concentration of $H_2O$ in iron melt is low and this effect on entropy decreases with increasing $X_{H_2O}^{Fe}$. Similar siderophile behaviors of $H_2O$ are also observed in the two-phase simulations, from which the calculated $D_{H_2O}^{Fe/MgSiO_3}$ are ~9 and 3 at 500 and 1000 GPa, respectively, in good agreement with the TI results (Extended Data Fig. 4).

Under given pressures, we find that $D_{H_2O}^{Fe/MgSiO_3}$ only marginally increases with decreasing temperature (Extended Data Fig. 3). For example, at $X_{H_2O}^{Fe} = 1.67$ wt%, $D_{H_2O}^{Fe/MgSiO_3}$ increases from 8.4 to 9.9 with temperature decreasing from 14200 K to 13000 K at 1000 GPa and the increase is even smaller (~1) at 500 GPa with temperature decreasing from 9800 K to 8600 K. We combined our data with previous lower-pressure results[9] and parameterized $D_{H_2O}^{Fe/MgSiO_3}$ as a function of $X_{H_2O}^{Fe}$ and pressure ($P$) (Fig. 1b). The temperature dependence was taken into account by correcting our calculated $D_{H_2O}^{Fe/MgSiO_3}$ to the values at the target temperatures, which are determined by the solidus of silicates[15,16] and melting curve of iron[17]. We find that $D_{H_2O}^{Fe/MgSiO_3}$ always increases with growing pressure, yet the rate of this increase declines rapidly (Fig. 1b). At pressures beyond 400 GPa, the increase becomes nearly imperceptible.

To quantify the effect of water partitioning on planet interiors, we consider four different interior scenarios (Fig. 2):

(A) rocky interior with no melting, and a separate water layer;
(B) rocky interior with (dry) melting in both mantle and core, and no water dissolution;
(C) same as B except with water dissolution in mantle melts;
(D) same as C except with water dissolution in core.

Differences between scenarios A, B, and C have been explored in detail in a previous study[8]. Characteristically, calculated radii are smallest when water is stored in both mantle and core (scenario D, Fig.2 & Fig. 3a), as a consequence of replacing water from the surface to fit within the iron and silicate structure. The addition of water reduces the density of both silicates and iron (Extended Data Fig. 5). Storing water deep in the interior implies that less water is available at the surface to act as a greenhouse gas, which lowers the surface temperature and hence the temperature of the interior. Meanwhile, water depresses the melting point. The interplay of these two effects of water results in deep magma oceans was observed in scenarios C and D (Extended Data Fig. 5). Fig. 3 shows that the relative difference between scenario D and others depends on bulk water content and planetary mass. For a bulk water mass fraction (wmf) of 3 wt%, we observe a relative difference in planet radius of up to 25%, while wmf of 20 wt% and 50 wt% exhibit differences of up to 14% (Fig. 3b). These differences are drastic compared to the observational uncertainties of well-characterized planets (down to a few percent in density). We note that the case of 50% bulk water mass fraction is an extreme end-member.

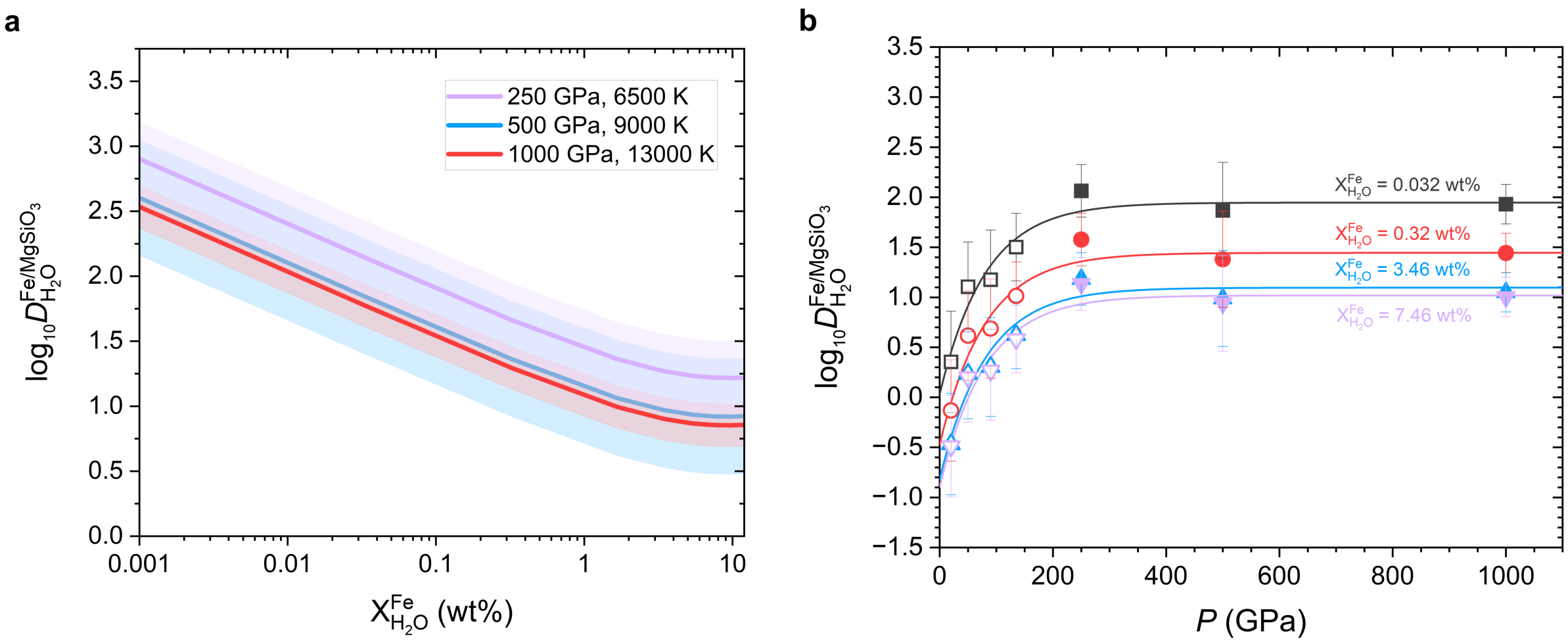

**Fig. 1. Partition coefficients of $H_2O$.** Calculated partition coefficients of $H_2O$ between iron and silicate melts ($D_{H_2O}^{Fe/MgSiO_3}$) as a function of its concentrations in iron melt ($X_{H_2O}^{Fe}$) **(a)** and $D_{H_2O}^{Fe/MgSiO_3}$ at various concentrations ($X_{H_2O}^{Fe}$) as a function of pressure ($P$) **(b).** Partition coefficients are calculated as the ratio of weight fractions of water in metal and silicate. **a,** The errors of $D_{H_2O}^{Fe/MgSiO_3}$ are represented by the shaded area with the same color as those of the respective lines. **b,** Open symbols are previous computational data[9]. Solid symbols represent our results. $D_{H_2O}^{Fe/MgSiO_3}$ is parametrized as a function of $H_2O$ concentration in iron melt, $X_{H_2O}^{Fe}$ (in wt%) and $P$ (in GPa) with the equation: $log_{10}\,D_{H_2O}^{Fe/MgSiO_3} = 1.26 - 0.34log_{10}X_{H_2O}^{Fe} + 0.08(log_{10}X_{H_2O}^{Fe})^2 - 4007/(1 + exp((P + 616)/80.6))$.

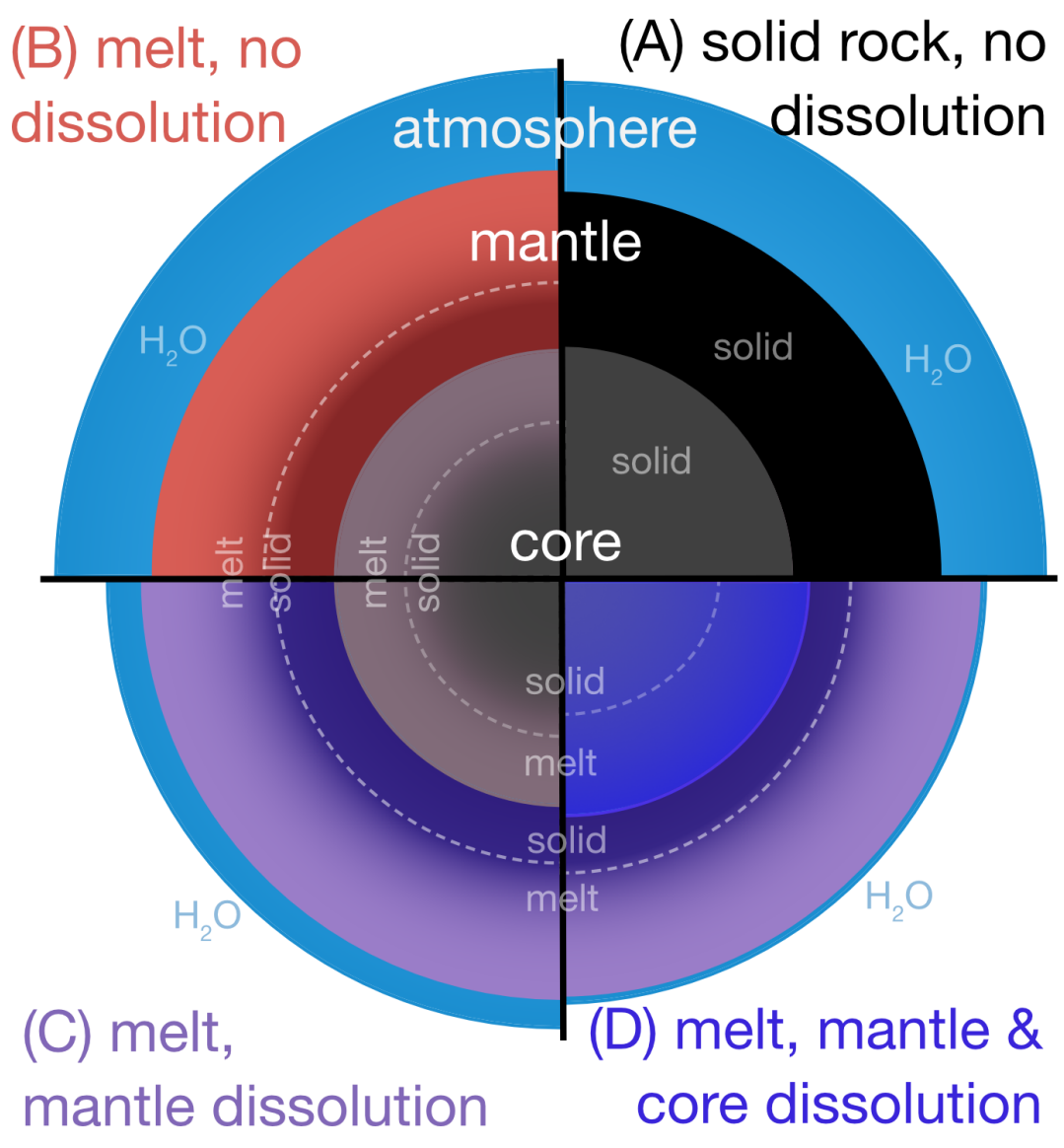

**Fig. 2. Four model scenarios employed in our study.** Model (A) closely resembles models presented in refs.[18,19] and other commonly used exoplanet interior models follow model (A) where silicate melts are neglected. Model (D) aligns closely with our current understanding of mineral physics and exoplanet interiors and is, therefore, the most accurate representation based on our existing knowledge.

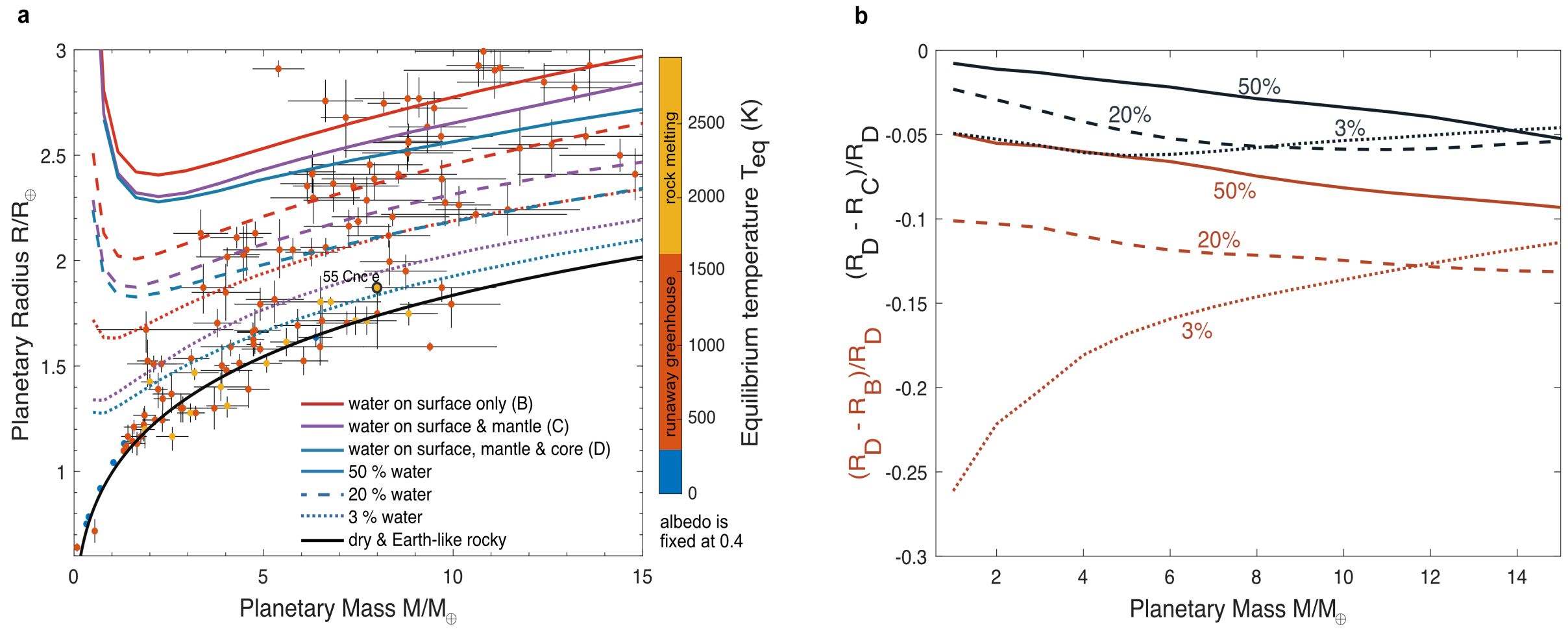

**Fig. 3. Effect of water partitioning into the core and mantle on mass-radius relationships: (a)** Mass-radius relation of exoplanets under various bulk water content and interior models (black: D versus C; red: D versus B following Fig. 2). Data for well-characterized exoplanets are shown for comparison. **(b)** Differences in calculated planet radii as a function of bulk water content, interior model, and planetary mass. Results are shown relative to the scenario D, the most self-consistent interior model. The rocky interior is composed of Mg-Si oxides and silicates (66 wt%) and an iron core of 33 wt% (see black line for a dry interior). We add different amounts of water (3%, 20%, and 50% of the total mass fraction). All curves are shown for an equilibrium temperature of 1000 K. The planet sample comes from the PlanetS database[20] (as of June 2023) to which we have added TOI-561 b, c, d[21] and the new estimates for Kepler-138 d, c, e[22]. We show planets with mass uncertainties better than 35% and radius uncertainties better than 30%. Differences between the model scenarios range from 3% to 25%. When neglecting water dissolution in the mantle and core, the errors are the largest. The largest errors are identified for small planet masses and lower water mass fractions, similar to what has been shown for the water dissolution in the mantle [8].

Planetary cores generally occupy a smaller volume than planetary mantles, at least for Earth-like rocky compositions and the generally anticipated range of rocky interiors[23]. Volumetrically speaking, the core is a smaller water reservoir than the mantle. However, iron can accommodate up to 70 times more water than silicates, depending on pressure and bulk water content. The net effect is that water is predominantly stored in the mantle for small-mass planets (below ~6 $M_\oplus$) and predominantly stored in the core for more massive planets (Fig. 4). Hence, for the majority of observed exoplanets most of bulk planetary water is stored deep in their cores. For example, a 9 $M_\oplus$ planet at $T_{eq}$ = 1000 K with a bulk water mass fraction of 50% will host 56% of the total water in the core, 41% in the mantle and only the remaining 3% on the surface. As the partitioning of water between iron and silicates is pressure-sensitive, the distribution of water in planets is closely linked to the planet's mass. For a hot Earth-mass planet ($T_{eq}$ = 700 K) with a 75 times greater water inventory than the present-day surface water on Earth (0.023%), we find that 58.4% of total water resides in the core, 40.1% in the mantle and 1.4% on the surface which corresponds to 0.024% of the total mass. This value is similar to the actual estimate of 0.023% water on Earth's surface. However, note that outgassing/ingassing and atmospheric loss to space can alter the initial surface water reservoir over the geological time.

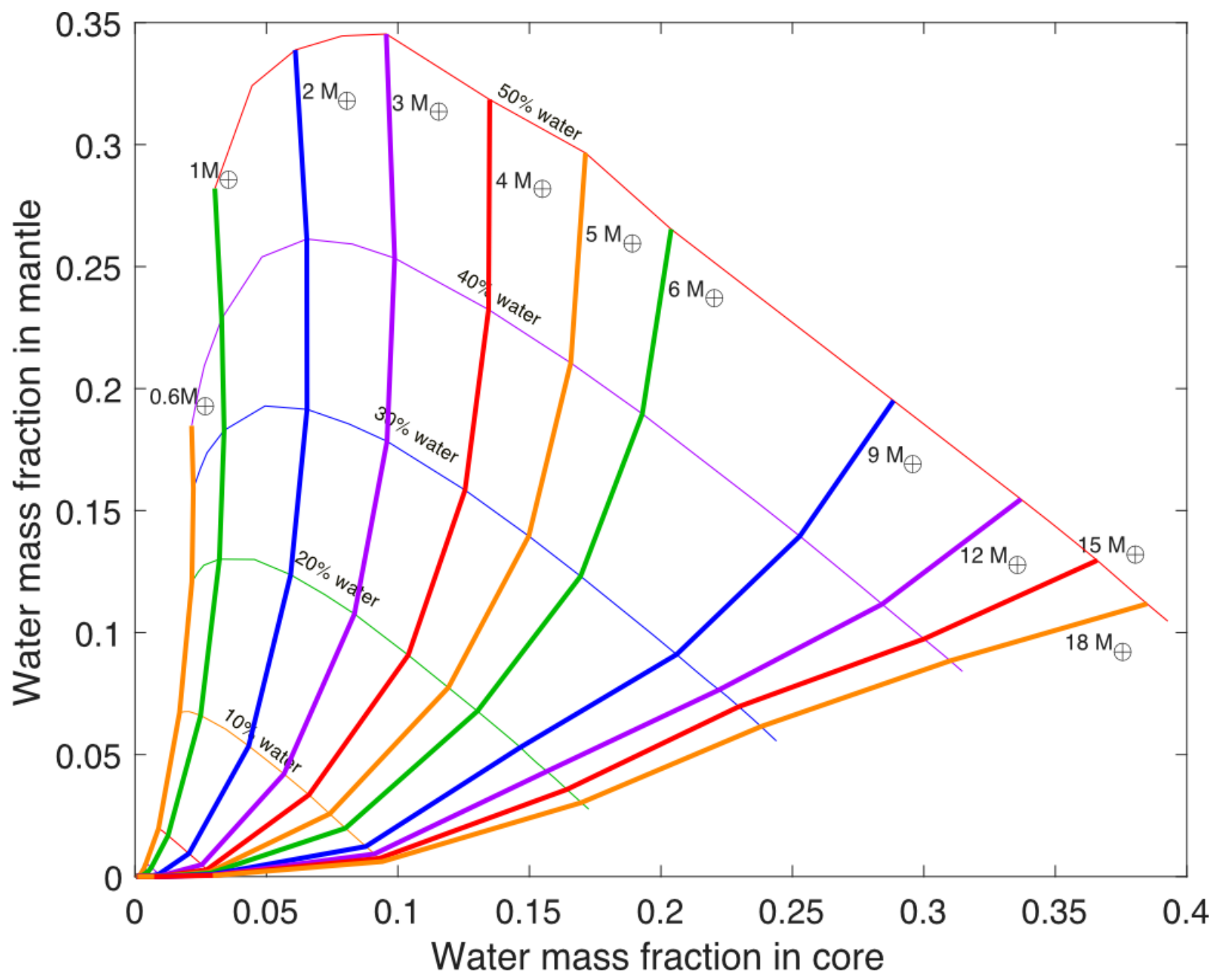

**Fig. 4. Water mass fractions stored in the mantle and the core depending on planet mass (thick lines) and bulk water mass fraction (thin lines).** As the partitioning of water is pressure dependent, the distribution of water within planets depends strongly on planet mass. The residual amount of water that is not stored in either mantle or core is kept on the surface. All lines have been calculated assuming an equilibrium temperature of 1000 K, and we note that the effect of temperature is negligible for the planets of interest.

Our interior model D is suited for warm and hot worlds inside the runaway greenhouse transition, where the presence of global magma oceans is expected. This scenario is particularly relevant for the majority of super-Earths and sub-Neptunes[8,24]. Compared to previous interior interpretations of observed exoplanets, our results from model D imply that planets may be more water-rich than previously thought, as commonly used mass-radius relationships overestimate the total radius for a given bulk water mass fraction (scenarios A and B). As a result, inferred water mass budgets from observed mass-radius data can differ significantly, with potential variations of up to nearly an order of magnitude (e.g., as seen in the overlap of red and blue curves for 3% and 20% wmf in Fig. 3a). Overall, our results on water partitioning between iron and silicate melt have fundamental implications for the distribution of water within planets, their densities and thermal structures, surface conditions, and ultimately their mass-radius relations.

## The case of 55 Cnc e

The enigmatic nature of 55 Cnc e, an extremely irradiated super-Earth with an unusually low density ($M_p = 8.59 \pm 0.43 M_\oplus$ , $T_{eq}$ = ~2000 K)[25] continues to captivate the community. Various models have been proposed to account for its large radius Rp = 1.947 +/– 0.038 $R_\oplus$ in relation to its mass, including nitrogen-rich, hydrogen-rich, or thin-metal-rich atmospheres on top of a magma oceans[26,27] or exotic rocky interiors depleted in iron and enriched in Na and Ca[28]. Water, if present on the surface, is expected to be in a state of loss. Observations in Ly-alpha of any extending atmosphere do not show any signals[29], which puts a low likelihood on the presence of surface water, although this could not be ruled out. In an extreme case, if 55 Cnc e lost all its mantle and surface water, but still holds water deep in its core, we can fit the mass-radius data with a water mass fraction of 34%, resulting in a radius of 1.946 $R_\oplus$ for a mass of Mp = 8.59, assuming an Earth-like core-mantle ratio. This amount corresponds to an initial bulk water budget of 55 Cnc e of much more than the usually assumed upper limit of 50% just after the planet's formation (Fig. 4). If we assume the equilibration between silicates and iron to happen at higher pressures, e.g., the pressure at the core-mantle boundary, this initial bulk water budget decreases to 45%. If this scenario holds true, it would indicate that the planet formed from water-rich building blocks originating from beyond the ice-line. Further investigations with evolution models are necessary to address whether this low-density planet can have water deep in its interior and whether this scenario requires a tenuous amount of surface water to still be exposed at the surface. In comparison to Dorn & Lichtenberg (2021)[8], we find that the majority of 55 Cnc e must be in the core and not in the mantle. This is an important difference, as water can be outgassed from the mantle, while it is locked up in the core for its lifetime.

**Sub-Neptunes**

The nature of sub-Neptunes is debated. Given mass and radius, the interiors of sub-Neptunes are degenerate. Hydrogen-dominated envelopes and water-rich interiors are discussed[30]. There are significant drawbacks despite the fact that water-rich interiors can fit their densities[31] and close-in water-rich worlds are predicted in planetary formation studies[32]. Mainly, the water world scenario does not explain two key observed correlations: radius distribution with orbital period and the bulk densities with stellar irradiation[30]. The presence of hydrogen-dominated envelopes remains the best explanation for their interiors as the two key observed correlations can be matched. In light of this, we compare our resulting mass-radius curves to the observed sub-Neptune population. Planets above 2.5 $R_\oplus$ require the addition of a hydrogen (Fig. 3) unless they could be more water-rich than 50%, which is difficult. Our 50% bulk water curve levels out around 2.5 $R_\oplus$ for higher masses and a 1000 K equilibrium temperature. As the majority of sub-Neptunes have equilibrium temperatures below 1000 K, it is therefore a conservative estimate. For example, GJ 1214 b[33], is unlikely to be a water world as previously suggested. Its radius of 2.742 $R_\oplus$ is about 7 sigma higher than that of the 50% water world scenario with 2.4 $R_\oplus$ at its equilibrium temperature of 596 K. We can expect hydrogen in its atmosphere which is advantageous for JWST observations as it increases the atmospheric scale height.

**Water-rich planets**

Recently, a class of water-rich planets around M-dwarfs has been proposed[34]. The six planets considered roughly follows a curve calculated for a composition of 50% water and 50% silicates (for $T_{eq}$ = 300 K). As the equilibrium temperatures of the six planets vary between 387 K and 1069 K, and the extent of water atmospheres is highly temperature-dependent within this range, we calculated different mass-radius relationships with scenario D. Clearly, the planet sample does not follow a single mass-radius curve (Fig. 5), indicating diversity among these exoplanets, likely with bulk water mass fractions ranging from 10%-40 %. All six planets have densities higher than the densities of planets with 50% of water. Besides water, the radii may also partly be explained by moderate-sized hydrogen-dominated envelopes. However, hydrogen-dominated atmospheres might only be possible in a narrow parameter space, especially on super-Earths smaller than ~7 $M_\oplus$[35].

Our results have significant implications for the potential habitability of water-rich worlds. Previous studies have suggested that super-Earths with too much water are unlikely to develop habitable surface conditions[36], as a solid layer of water may separate liquid water from silicate minerals in the mantle[37,38]. However, our findings imply that worlds with thick water layers may not be common, as the majority of water on super-Earths will likely be locked in the core instead of the surface, especially for more massive planets (Fig. 4). Hence, even water-rich planets with several tens of water mass fractions may have the potential to develop Earth-like surface conditions. Our research thus opens up new possibilities for understanding the habitability of exoplanets and sheds light on the potential existence of water-rich worlds that can support life.

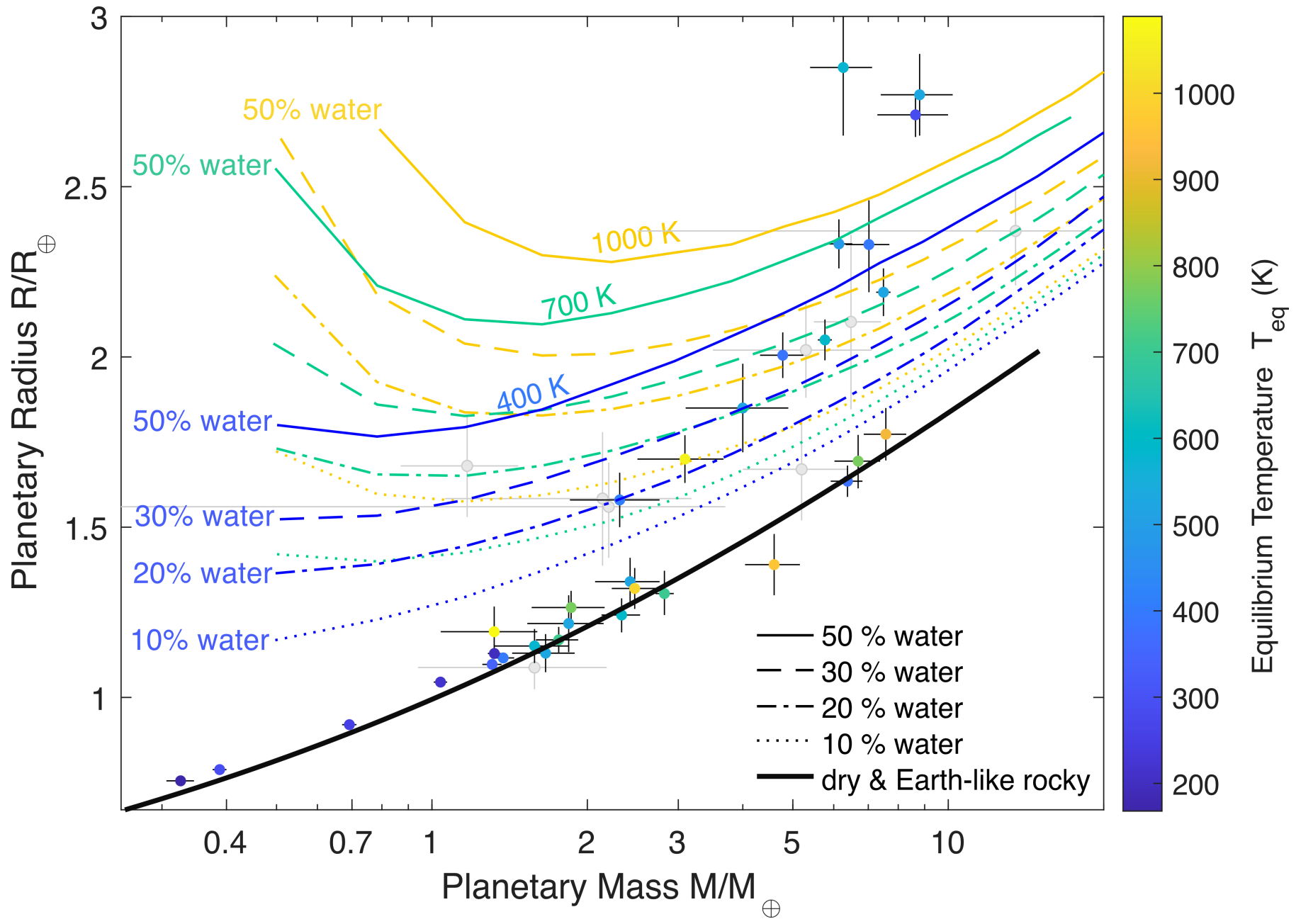

**Fig. 5. Mass-radius curves compared with the sample of small transiting planets around M dwarfs from Luque and Pallé (2022)[34].** All curves are calculated using scenario D, if not mentioned otherwise. Colored curves are calculated for different equilibrium temperatures and different water mass fractions (see line styles). The curves at 1000 K and 3%, 20%, 50% water mass fractions are also shown in Fig. 3, as is the solid black line. The planets previously interpreted to be water-rich (~50%)[34] are likely to be more diverse in composition and contain less than 50% of water.

**Acknowledgements:** The work reported in this paper was conducted using the Princeton Research Computing which is a consortium of groups led by the Princeton Institute for Computational Science and Engineering (PICSciE) and the Office of Information Technology's Research Computing.

**Funding:** J. D. acknowledges support from the National Science Foundation under Grant No. EAR-2242946. C.D. acknowledges support from the Swiss National Science Foundation under grant TMSGI2_211313. In parts, this work has been carried out within the framework of the NCCR PlanetS supported by the Swiss National Science Foundation under grants 51NF40_182901 and 51NF40_205606.

**Author contributions**

Conceptualization: J.D. and C.D.
Molecular dyanmics simulation: H.L.
Planetary interior modelling: C.D.
Data analysis and interpretation: H.L., C.D., and J.D.
Writing-original draft: H.L. and C.D.
Writing-review and editing: H.L., C.D., and J.D.

**Competing interests:** The authors declare that they have no competing interests.

**Data and materials availability:** All data are available in the main text or the supplementary materials. Raw molecular dynamics simulations will be publicly available from Figshare. Any other raw simulation outputs are available from the corresponding author on request.

## Methods

### Calculations of chemical potentials and partition coefficients

The method of calculating the chemical potential of $H_2O$ in iron (Fe)/silicate ($MgSiO_3$) melt is outlined in detail in previous studies[9,13] and are briefly summarized here. The Gibbs free energy of a system at specific pressure (*P*), temperature (*T*), and solute concentration (*x*) can be written as

$$G(p,T,x) = \bar{G}(p,T,x) - TS_{mix}^{\text{system}} \quad (1)$$

where $\bar{G}(p,T,x)$ is the excess free energy and $S_{mix}^{\text{system}}$ is the ideal gas mixing entropy. Then the chemical potential of $H_2O$ can be expressed as

$$\mu_{\text{H}_2\text{O}}(p,T,x) = \bar{\mu}_{\text{H}_2\text{O}}(p,T,x) - TS_{mix}^{\text{H}_2\text{O}} \quad (2)$$

Here $\bar{\mu}_{\text{H}_2\text{O}}(p,T,x)$ and $S_{mix}^{\text{H}_2\text{O}}$ can be calculated by

$$\bar{\mu}_{\text{H}_2\text{O}}(p,T,x) = \bar{G}(p,T,x) + (1-x)\frac{\partial \bar{G}(p,T,x)}{\partial x} \quad (3)$$

$$S_{mix}^{\text{H}_2\text{O}} = S_{mix}^{\text{system}} + (1-x)\frac{\partial S_{mix}^{\text{system}}}{\partial x} \quad (4)$$

The variations of $\bar{G}(p,T,x)$ per formula with respect to $x$ for our systems, $\text{Fe}_{1-\text{x}}(\text{H}_2\text{O})_\text{x}$ and $(\text{MgSiO}_3)_{1-\text{x}}(\text{H}_2\text{O})_\text{x}$, were found to be fitted well using a linear equation:

$$\bar{G}(p,T,x) = a + bx \quad (5)$$

where a and b are the fitting parameters. As a result, $\bar{\mu}_{\text{H}_2\text{O}}(p,T,x)$ can be calculated as $a+b$ and the errors of $\bar{\mu}_{\text{H}_2\text{O}}(p,T,x)$ can be derived using error propagation based on the uncertainties of $a$ and $b$.

The $S_{mix}^{\text{H}_2\text{O}}$ for the four systems calculated using the equation (4) are:

$$S_{mix}^{\text{H}_2\text{O in Fe}} = -k_B\left(2\ln\frac{2x}{1+2x} + \ln\frac{x}{1+2x}\right) \quad (6)$$

$$S_{mix}^{\text{H}_2\text{O in MgSiO}_3} = -k_B\left(2\ln\frac{2x}{5-2x} + \ln\frac{3-2x}{5-2x}\right) \quad (7)$$

where $k_B$ is the Boltzmann constant. Then at equilibrium when $\mu_{\text{H}_2\text{O}}^{\text{Fe}}(p,T,x) = \mu_{\text{H}_2\text{O}}^{\text{MgSiO}_3}(p,T,y)$, the partition coefficients $D_{\text{H}_2\text{O}}^{\text{Fe/MgSiO}_3} = x/y$, can be easily derived using the equation (2).

### Molecular dynamics simulations

We simulated iron melts and silicate melts using supercells consisting of 64 Fe atoms and 32 formulae of $MgSiO_3$, respectively. $H_2O$ were later added to the supercells for water-bearing systems. Born-Oppenheimer *ab initio* molecular dynamics (AIMD) simulations were conducted using the projector augmented wave method[39,40] and PEBsol exchange-correlation functional[41], as implemented in VASP[42]. The valence configurations we used for elements are Fe $3p^63d^74s^1$, Mg $2p^63s^2$, Si $3s^23p^2$, O $2s^22p^4$, and H $1s^1$. Ions and electrons were assumed to be in thermal equilibrium via the Mermin functional[43]. We ran each simulation up to 20,000 steps with a timestep of 0.5 femtosecond. Gamma Point Brillouin zone sampling and a plane wave energy cutoff of 600 eV and 800 eV for iron and silicate melts, respectively, were initially utilized. We then corrected the pressure for iron by employing 2×2×2 Monkhorst-Pack mesh of k-points, as the system size we used for iron is small. A canonical ensemble (*NVT*) with temperature controlled by a Nosé-Hoover thermostat was adopted[44]. To determine the equilibrium volume, we performed several *NVT* simulations using different volumes around the target pressure and then fitted the pressure-volume data. Radial distribution function (RDF) and mean-square displacement (MSD) were examined to make sure that our simulated systems were in the liquid state.

### Thermodynamic integration

We calculated the Gibbs free energy of iron and silicate melts by thermodynamic integration from ideal gas[45]. To do this the difference in internal energy between the melt ($U_1$) and the ideal gas ($U_0$) with respect to the coupling parameter λ is integrated by

$$\Delta G_{0-1} = \int_0^1 d\lambda \langle U_1(R) - U_0(R) \rangle_\lambda \quad (10)$$

To overcome the numerical instabilities resulting from close contacts between non-interacting atoms at $\lambda = 0$, we employed the variable transformation approach[28] with $\lambda(x) = \left(\frac{x+1}{2}\right)^{1/(1-k)}$, which yields

$$\Delta G = \frac{1}{2(1-\kappa)} \int_{-1}^{1} f(\lambda(x))\, \lambda(x)^k dx \quad (11)$$

where we adopted $k = 0.8$. We used an 8-point Gauss-Lobatto rule to evaluate this integral. The errors of internal energies were calculated by blocking average method[46]. We obtained the errors of the numerically integrated free energies by error propagation.

The Gibbs free energy of the ideal gas was calculated as

$$G^{ig} = F^{ig} + PV = -k_B T \sum_i ln\left(\frac{V^{N_i}}{N_i! \Lambda_i^{3N_i}}\right) + PV \quad (12)$$

where $k_B$ being the Boltzmann constant, $T$ temperature, $P$ pressure, $V$ volume, $N_i$ is the number of atoms for element $i$, $\Lambda_i$ is the thermal wavelength of element $i$ given by $\Lambda_i = h/(2\pi M_i k_B T)^{1/2}$, with $h$ the Plank's constant, $M_i$ the atomic mass of element $i$. We calibrated all the calculations to the target pressures using the thermodynamic relation:

$$G(P_2, T) - G(P_1, T) = \left(\int_{P_1}^{P_2} V dP\right)_T \quad (13)$$

**Planet interior calculations**

We employ a 1-D interior model that describes a planet in hydrostatic equilibrium which incorporates three components: a core, a mantle and a water layer (of steam, liquid, solid, or supercritical). We solve the equation of mass conservation, the equation of hydrostatic equilibrium, the equation of thermal transport, and the equation of states for different materials/phases. Our model is built upon Dorn and Lichtenberg (2021)[8], with major improvements that include the equation of states for dry and wet iron melts and the dissolution of water in the core. For the scope of this study, we exclude an atmosphere of other high mean molecular weight species or H/He, and the only volatile considered is $H_2O$.

**The Core**

We consider a core made of Fe, H, and O. For solid Fe, we use the equations of state for hexagonal close packed (hcp) iron at $P >= 310$ GPa[47] and at $P < 310$ GPa[48]. For liquid iron and liquid iron with $H_2O$, we perform *ab initio* molecular dynamics simulations and calculate their equations of state (*PVT*) under the extreme pressure and temperature conditions relevant to Super-Earths and sub-Neptunes (~8000 K–14000 K & ~50 GPa–1300 GPa). The Mie–Grüneisen equation of state for dry liquid iron (64 Fe atoms) and hydrous liquid iron (64 Fe atoms and 8 $H_2O$) can be expressed as:

$$P = \frac{3}{2} K_{T_0} \left[\left(\frac{V_0}{V}\right)^{\frac{7}{3}} - \left(\frac{V_0}{V}\right)^{\frac{5}{3}}\right] \left[1 + \frac{3}{4}(K'_{T_0} - 4)\left\{\left(\frac{V_0}{V}\right)^{\frac{2}{3}} - 1\right\}\right] + \frac{(T-T_0)}{1000}\left[a + b\left(\frac{V_0}{V}\right) + c\left(\frac{V_0}{V}\right)^2\right] \quad (14)$$

where, for dry liquid iron, $K_{T_0} = 49.249$ GPa, $K'_{T_0} = 4.976$, $V_0 = 1043.912$ Å$^3$, $T_0 = 8000$ K, $a = -15.957$, $b = 20.946$, $c = -3.811$. For hydrous liquid iron, $K_{T_0} = 57.162$ GPa, $K'_{T_0} = 4.535$, $V_0 = 1181.389$ Å$^3$, $T_0 = 8000$ K, $a = -8.478$, $b = 12.049$, $c = -1.258$. The phase transitions are calculated following Anzellini et al. (2013)[49]. The melting curve is calculated following recent experimental[17] and computational[50] works by assuming a linear H and O concentration dependence: $T_\mathrm{m} = T_\mathrm{m_hcp_Fe} - a_1 x - b_1 y$, where $a_1 = 1031$ K and $b_1 = 100$ K, and $T_\mathrm{m_hcp_Fe} = 5530 \times ((P - 260)/293 + 1)^{0.552}$. The parameter $x$ is the mass fraction of H in wt% and $y$ the mass fraction of O in wt%.

The core thermal profile is assumed to be adiabatic throughout the core. At the core-mantle boundary (CMB), there is a temperature jump as the core can be hotter than the mantle due to the residual heat released during core formation. Following Stixrude (2014)[15], this temperature jump depends on the melting temperature of the silicate mantle. If the initially calculated temperature at the CMB is less than the melting temperature of the mantle material, the CMB temperature is increased up to the melting temperature.s

**The Mantle**

The calculation of the mantle structure is described in detail in Dorn and Lichtenberg (2021)[8] , with few adaptations. In brief, the mantle is assumed to be made up of three major components, i.e., MgO, $SiO_2$, FeO. For the solid mantle, we use the thermodynamical model Perple_X[51], to compute stable mineralogy and density for a given composition, pressure, and temperature, employing the database of Stixrude and Lithgow-Bertelloni (2022)[52]. For pressures higher than ~125 GPa, we define stable minerals a priori and use their respective equation of states from various sources[52-55]. We consider $SiO_2$ from Faik et al. (2018)[55], post-perovskite (Mg, Fe)$SiO_3$ from Hemley et al. (1992)[56], FeO from Fischer et al. (2011)[53] and MgO from Musella et al. (2019)[54]. For the liquid mantle, we calculate its density assuming a ideal mixture of components as in Dorn and Lichtenberg (2021)[8]. Specifically, we use the equations of state for $Mg_2SiO_4$ from Stewart et al. (2020)[57], for $SiO_2$ at $P$ >= 20 GPa from Faik et al. (2018)[55], for $SiO_2$ at $P$< 20 GPa from Melosh (2007)[58], and for FeO from Ichikawa and Tsuchiya. (2020)[59]. To compute mixtures of the above components, we use the additive volume law[57]. We use $Mg_2SiO_4$ instead of MgO since the data for forsterite has been recently updated for the high pressure temperature regime, which is not available for MgO to our knowledge.

The addition of water reduces the density, for which we follow Bajgain et al. (2015)[60] and reduce the silicate density by 0.036 g/cm$^3$ per wt% water. For small water mass fractions, this reduction is nearly independent of pressure and temperature. At large water mass fractions, a constant density reduction becomes invalid. Thus, we also calculate the density of the rock-water mixture with the additive volume law and use the maximum density of both values.

The melting curve is calculated following Belonoshko et al. (2005)[61] (P<189.75 GPa) and Stixrude (2014)[15] ($P$ >= 189.75 GPa). The melting temperature of pure $MgSiO_3$ ($T_{m,dry}$) changes by compositional changes, specifically the addition of water lowers the melting temperature, for which we follow Katz et al. (2003)[62]: $T_{\mathrm{m,wet}} = T_{\mathrm{m,dry}} - a_2 x_{\mathrm{H_2O}}{}^{b_2}$, where $a_2$ = 43 K and $b_2$= 0.75, and $x_{\mathrm{H_2O}}$ water mass fraction of the silicate melt. For additions of FeO into silicates we follow Dorn et al. (2018)[28].

**The water layer**

Water can be in solid, liquid, steam or supercritical phase. We use the equation of state (EoS) compilation of Haldemann et al. (2020)[63]. The transit radius of a planet is assumed to be at a pressure of $P_{\mathrm{Transit}}$ = 1 mbar. This is a simplification as the transit radius depends on temperature, while its effect on the planets of interest is small[64]. The thermal profile is assumed to be fully adiabatic, except for pressures smaller than the pressure at the tropopause which we fix at 0.1 bar, and where we keep an isothermal profile as in Dorn & Lichtenberg (2021)[8].

**Water-melt mixtures**

For the model scenarios C and D, we allow the water to be mixed into the mantle melt, and both mantle & core melts, respectively. The partitioning between mantle melts and the water layer is determined by a modified Henry's law, for which empirical data are input and summarized in Bower et al. (2022)[65] (< 300 bar), Lichtenberg et al. (2021)[7] (<1 GPa), and Kessel et al. (2005)[66] (for the second critical endpoint). We use a fitted solubility function, which can be found in Dorn & Lichtenberg (2021)[8]. For the partitioning of water between iron and silicates, we follow the dependencies as described in the main text (see also Fig. 1). For the equilibration pressure, we use half of the core-mantle boundary pressure.

Water plays a crucial role in melt processes. As described above, we account for the effect of water in reducing the melt temperature and melt density. In principle, water can also be taken up by solid rocks. However, the solubility of water in solid rocks is orders of magnitude lower than in melts. Thus, we neglect the effect of hydrated solid silicates on the total radius as it only accounts for a few percent at maximum and only in extreme cases[67].

## Oxidation of iron

$H_2O$ may act as an oxidizing agent in the planetary interior. The possibility of fully oxidizing metallic iron by water to form coreless exoplanets was discussed[68] while explicitly assuming that water oxidizes iron. Indeed, FeO is expected in water-rich planetesimals in the form of clays at low temperature[69]. However, these FeO-rich clays destabilize at higher temperatures before melting occurs (beyond ~1000 K)[70]. During magma ocean stage, the extent of iron oxidation involves a dynamic interplay between the rate of oxidation, the rate at which the metallic iron descends, the size of the metal drop, and the reactions between different species such as $H_2$, $H_2O$, Fe, and FeO. We show below that the effect of iron oxidation (if any) on the mass-radius relationship is relatively small. In fact, iron oxidation would lead to a further decrease in the radius compared to the scenario D we consider.

First of all, recently the thermodynamic modeling suggests that both water and metallic iron may be produced via the redox-reactions between primordial hydrogen and mantle melts[6]. More specifically, the reactions $FeO + H_2 <-> Fe + H_2O$ and $SiO_2 + H_2 <-> Si + H_2O$ will predominantly lean towards to right-hand side for super-Earths and sub-Neptunes, which suggest that (1) iron reduction, rather than oxidation, is dominant during the core formation and (2) that our calculations are relevant also for planets whose water originates from endogenic water in addition to primordially accreted water.

Then, it is found that the equations of state (*PVT*) of $Mg_{1-x}Fe_xSiO_3$ liquid is very similar to that of pure $MgSiO_3$ liquid[71]. Therefore, even if a big chunk of iron is oxidized, the presence of oxidized ferrous iron in the silicate melt should not have a significant influence on our calculated metal-silicate partition coefficients of water. While our calculations realistically quantify the water concentration in the core, the size and the mass of the core may be affected by iron oxidation. To investigate the effect of iron oxidation on calculated radii, we extrapolate the partitioning coefficients to the extreme case of 50% bulk water mass fraction. Because it remains unknown at the core formation conditions of super-Earths how much iron is oxidized given a bulk water mass fraction, we directly add different amounts of iron as mantle FeO for the 50% bulk water case and keep the total amount of oxygen constant (Extended Data Fig. 6). In cases where water oxidizes less than 50% of the iron as mantle FeO, the decrease in total radius is less than 2%. If water oxidizes more than 50% of the iron, the radius decrease is significant (up to 10%). We highlight that the case of 50% bulk water mass fraction is extreme and we include it as this end-member is commonly discussed in the literature[72].

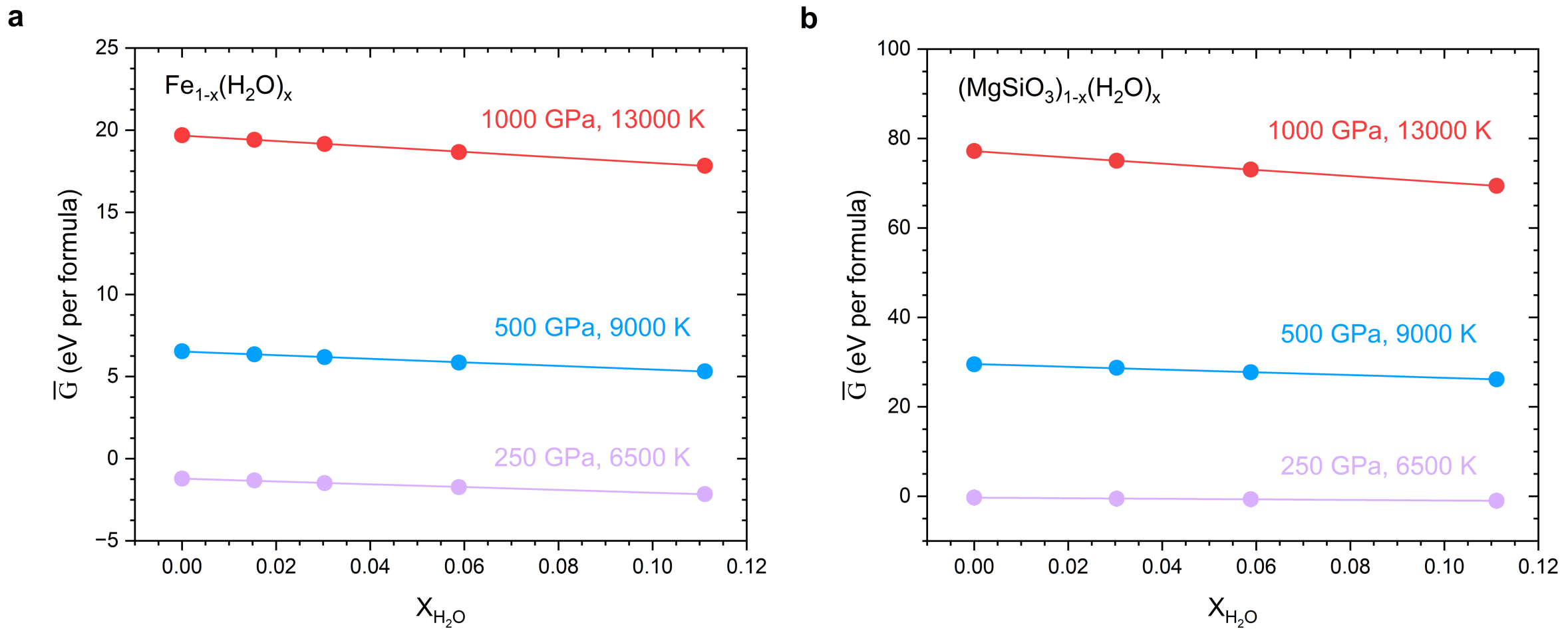

**Extended Data Fig. 1. Calculated Gibbs free energies of $\mathrm{Fe}_{1-x}(\mathrm{H_2O})_x$ and $(\mathrm{MgSiO_3})_{1-x}(\mathrm{H_2O})_x$ melts as a function of $H_2O$ mole fractions ($X_{\mathrm{H_2O}}$).**

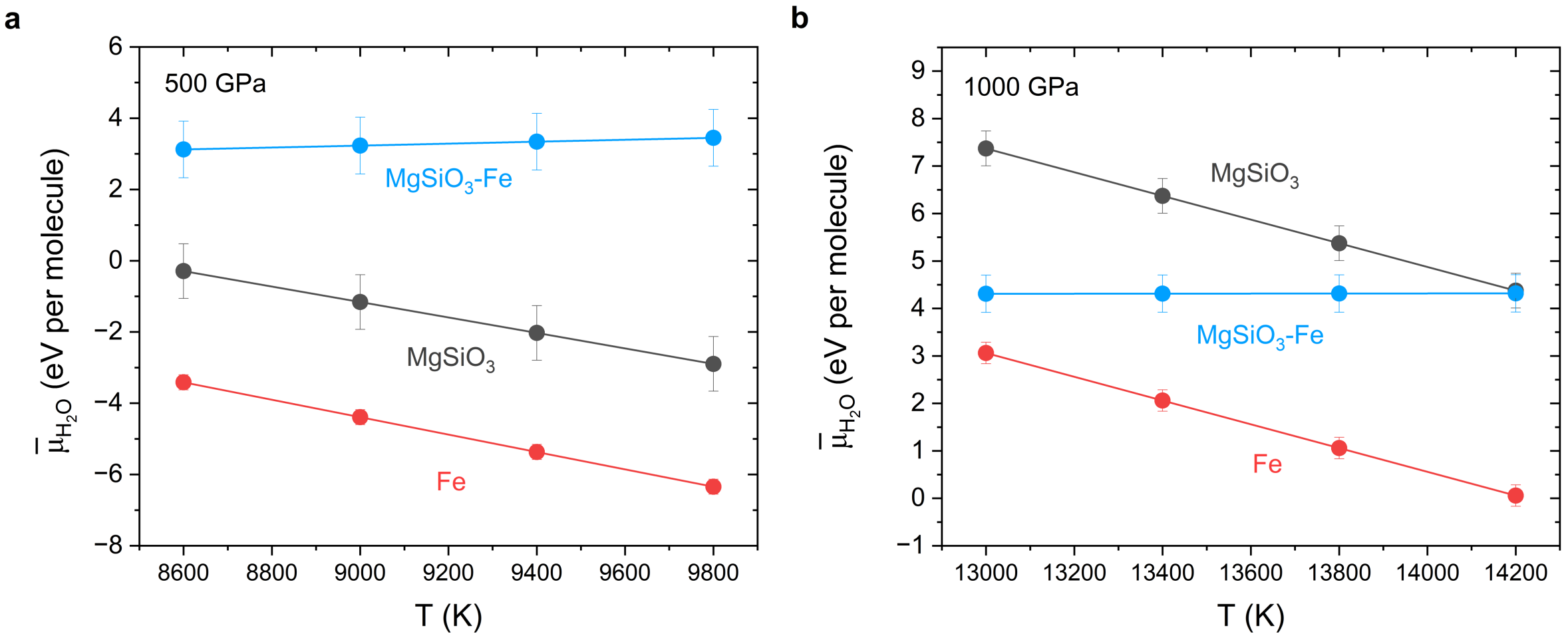

**Extended Data Fig. 2. Calculated chemical potentials of $H_2O$ in iron (Fe) and silicate ($MgSiO_3$) melts and their differences ($MgSiO_3$-Fe) at different temperatures.**

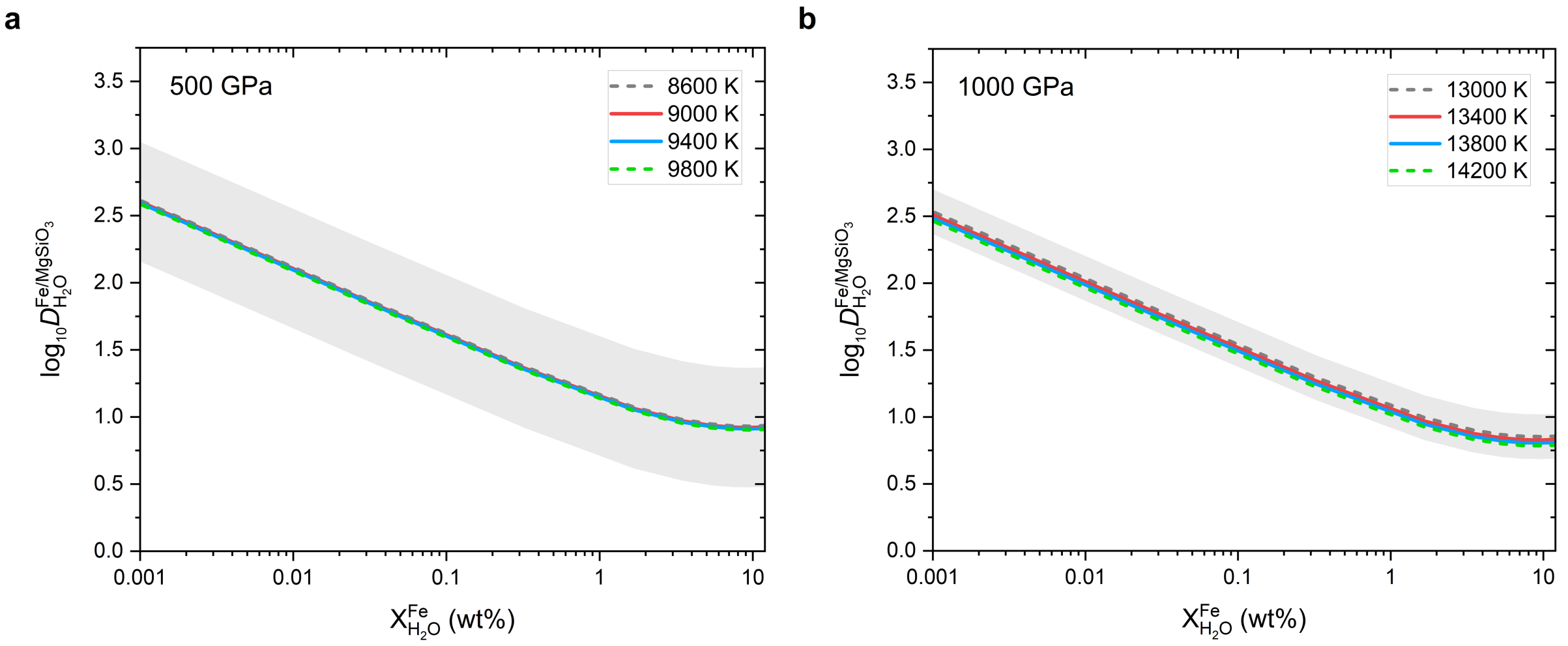

**Extended Data Fig. 3. Calculated partition coefficients of $H_2O$ between iron and silicate melts ($D_{H_2O}^{Fe/MgSiO_3}$) as a function of its concentrations in iron melt ($X_{H_2O}^{Fe}$) at different temperatures.**

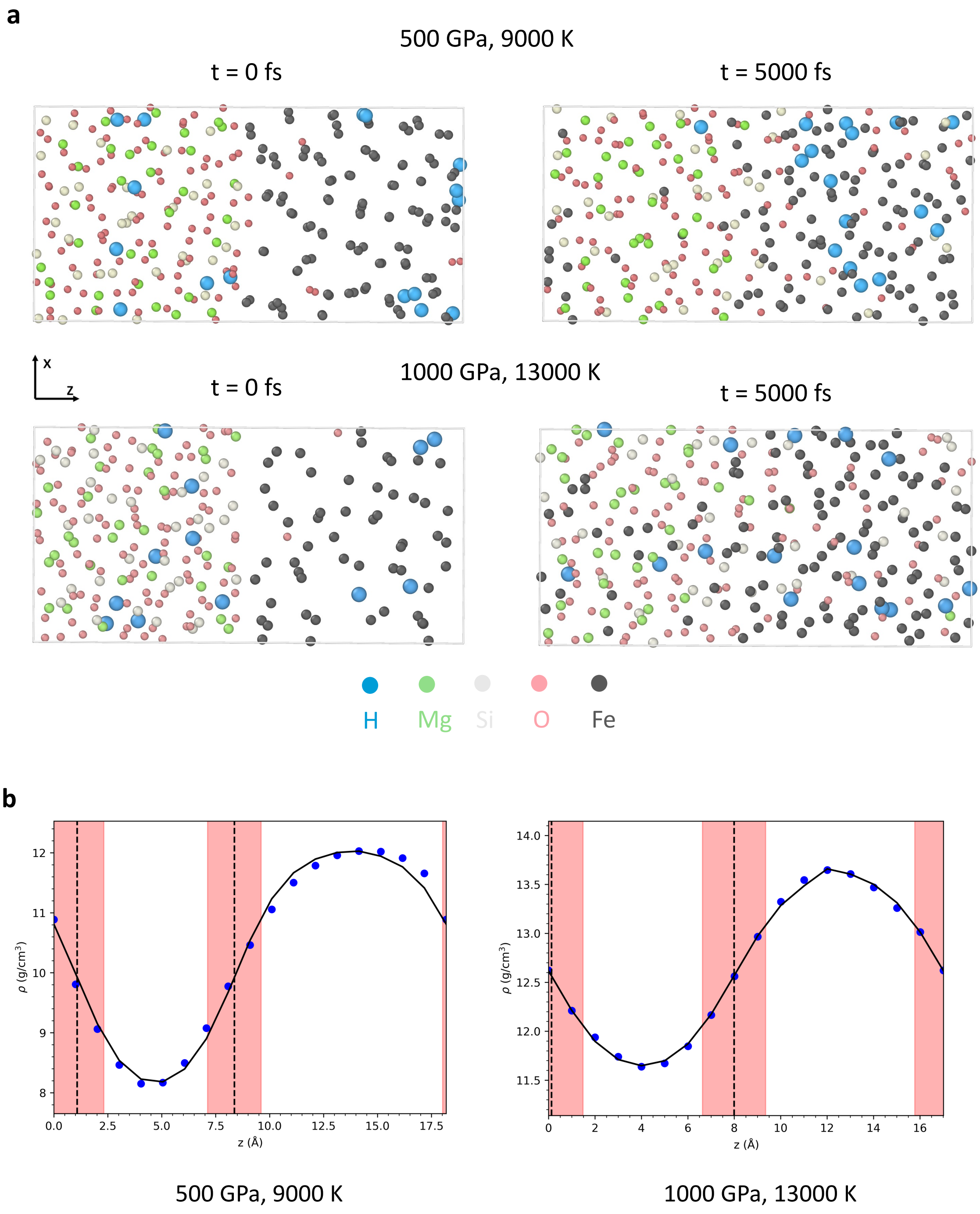

**Extended Data Fig. 4. Two-phase molecular dynamics simulations of water partitioning between iron (Fe) and silicate ($MgSiO_3$) melts at 500 GPa/9000 K and 1000 GPa/14000 K. a,** The initial configurations and the snapshots at 5000 femtoseconds (fs) are shown. b, The corresponding instantaneous coarse-grained density profile (blue dots) along the z-axis of the simulation box at 5000 fs and the best fitting curve (solid line). The dashed vertical lines represent the locations of Gibbs dividing surfaces.

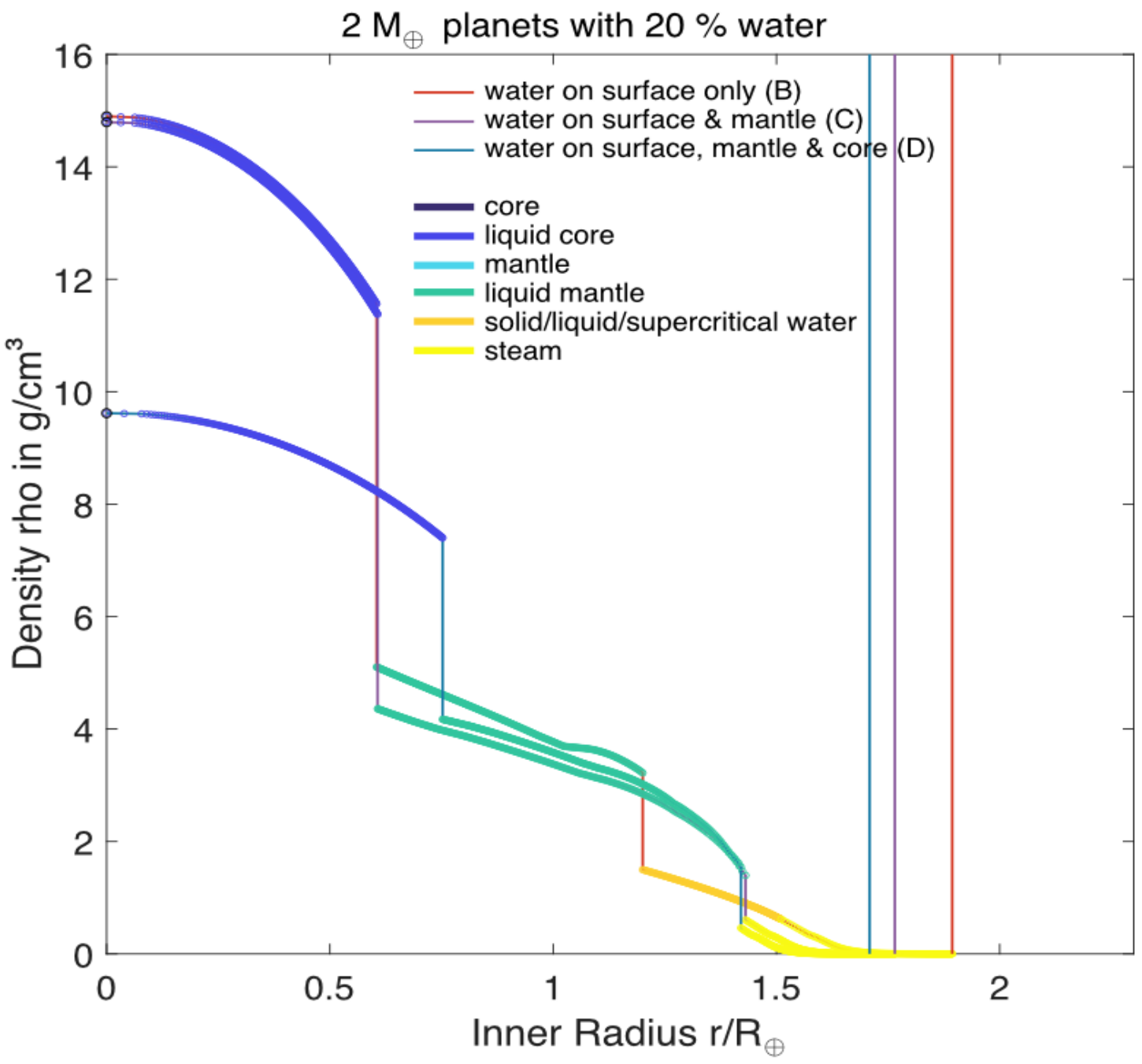

**Extended Data Fig. 5. Density profiles for planets of 2 M⊕ and 20 % bulk water mass fraction at $T_{eq}$ of 700 K.** The total radius for each of the scenarios is highlighted as a vertical line (red: B, lilac: C, blue: D). The different distributions of water in planets change density structure, thermal structure (not shown), melting temperatures (not shown), and hence the extent of molten layers. Wherever the density profiles seem to flatten, we checked that all densities increase at all pressures.

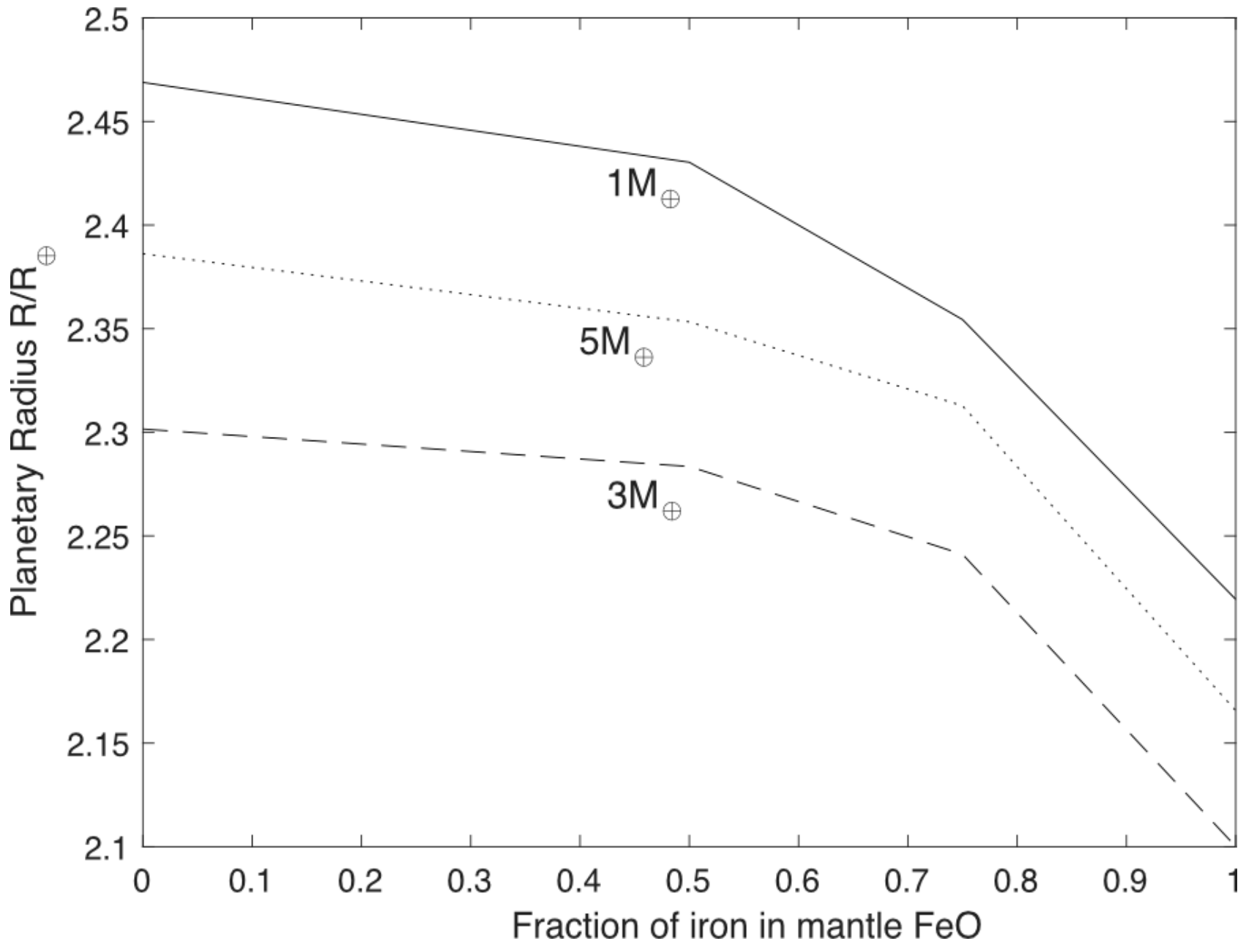

**Extended Data Fig. 6. The effect of iron oxidation on the radii for the 50% bulk water case at $T_{eq}$ of 1000 K.** We assume that different fractions of iron are added to the mantle by the oxidation of water. The scenario D as shown in Fig.3 assumes zero amounts of FeO in the mantle. Radii for 1 $M_{\oplus}$ planets are largest as in Fig.3 as steam atmospheres become less gravitational bound towards planets of low mass.

**Extended Data Table 1. Calculated free energies.** Simulations were conducted at 250, 500, and 1000 GPa, corresponding to temperatures of 6500, 9000, and 13000 K, respectively.

| P (GPa) | T (K) | System | V (Å$^3$/atom) | $\bar{G}$ (eV/atom) | System | V (Å$^3$/atom) | $\bar{G}$ (eV/atom) |
|---|---|---|---|---|---|---|---|
| 250 | 6500 | | 7.544 | –1.215±0.007 | | 5.694 | –0.058±0.006 |
| 500 | 9000 | $Fe_{64}$ | 6.434 | 6.538±0.012 | $(MgSiO_3)_{32}$ | 4.696 | 5.905±0.007 |
| 1000 | 13000 | | 5.353 | 19.691±0.017 | | 3.780 | 15.440±0.008 |
| 250 | 6500 | | 7.314 | –1.288±0.010 | | 5.641 | –0.107±0.006 |
| 500 | 9000 | $Fe_{64}(H_2O)_1$ | 6.232 | 6.160±0.014 | $(MgSiO_3)_{32}$ $(H_2O)_1$ | 4.646 | 5.815±0.008 |
| 1000 | 13000 | | 5.189 | 18.836±0.016 | | 3.744 | 15.189±0.008 |
| 250 | 6500 | | 7.108 | –1.403±0.008 | | 5.593 | –0.137±0.005 |
| 500 | 9000 | $Fe_{64}(H_2O)_2$ | 6.057 | 5.831±0.014 | $(MgSiO_3)_{32}$ $(H_2O)_2$ | 4.604 | 5.681±0.008 |
| 1000 | 13000 | | 5.035 | 18.060±0.016 | | 3.709 | 14.964±0.008 |
| 250 | 6500 | | 6.754 | –1.548±0.007 | | 5.491 | –0.211±0.006 |
| 500 | 9000 | $Fe_{64}(H_2O)_4$ | 5.738 | 5.237±0.010 | $(MgSiO_3)_{32}$ $(H_2O)_4$ | 4.522 | 5.472±0.007 |
| 1000 | 13000 | | 4.767 | 16.702±0.014 | | 3.642 | 14.531±0.007 |
| 250 | 6500 | | 6.189 | –1.765±0.007 | | | |
| 500 | 9000 | $Fe_{64}(H_2O)_8$ | 5.239 | 4.352±0.008 | | | |
| 1000 | 13000 | | 4.342 | 14.594±0.011 | | | |

**Extended Data Table 2. Calculated free energies as a function of temperature at 500 and 1000 GPa.**

| P (GPa) | T (K) | System | V (Å$^3$/atom) | $\bar{G}$ (eV/atom) | System | V (Å$^3$/atom) | $\bar{G}$ (eV/atom) |
|---|---|---|---|---|---|---|---|
| 500 | 8600 | $Fe_{64}$ | 6.407 | 7.019 | $(MgSiO_3)_{32}$ | 4.670 | 6.236 |
| | 9000 | | 6.434 | 6.538 | | 4.696 | 5.905 |
| | 9400 | | 6.452 | 6.039 | | 4.711 | 5.565 |
| | 9800 | | 6.465 | 5.537 | | 4.736 | 5.216 |
| 500 | 8600 | $Fe_{64}(H_2O)_4$ | 5.719 | 5.689 | $(MgSiO_3)_{32}$ $(H_2O)_2$ | 4.582 | 6.012 |
| | 9000 | | 5.738 | 5.237 | | 4.604 | 5.681 |
| | 9400 | | 5.762 | 4.762 | | 4.622 | 5.343 |
| | 9800 | | 5.776 | 4.287 | | 4.643 | 4.998 |
| 1000 | 13000 | $Fe_{64}$ | 5.353 | 19.691 | $(MgSiO_3)_{32}$ | 3.780 | 15.437 |
| | 13400 | | 5.366 | 19.160 | | 3.791 | 15.069 |
| | 13800 | | 5.378 | 18.636 | | 3.803 | 14.695 |
| | 14200 | | 5.387 | 18.105 | | 3.811 | 14.317 |
| 1000 | 13000 | $Fe_{64}(H_2O)_4$ | 4.767 | 16.702 | $(MgSiO_3)_{32}$ $(H_2O)_2$ | 3.709 | 14.964 |
| | 13400 | | 4.778 | 16.218 | | 3.719 | 14.599 |
| | 13800 | | 4.786 | 15.723 | | 3.726 | 14.227 |
| | 14200 | | 4.796 | 15.224 | | 3.737 | 13.849 |

# Supplementary Materials for

**Majority of water hides deep in the interiors of exoplanets**

Haiyang Luo[1*†], Caroline Dorn[2*†] and Jie Deng[1*]

[1]Department of Geosciences, Princeton University, Princeton, 08544 New Jersey, USA.
[2]Institute for Particle Physics and Astrophysics, ETH Zürich, Otto-Stern-Weg 5, 8093 Zürich, Switzerland.
*Corresponding authors. Email: haiyang.luo@princeton.edu, dornc@ethz.ch, or jie.deng@princeton.edu
[†]These two authors contribute equally to this work.

## The case of hydrogen

We calculate the metal-silicate partition coefficients of H at 500 and 1000 GPa (Supplementary Fig. 2). It is shown that H remains siderophile and exhibits similar partition coefficients as those of $H_2O$ at low concentration (e.g., $X_{\mathrm{H_2O}}^{\mathrm{Fe}} = 0.032\ \mathrm{wt\%}$). We mainly consider $H_2O$ instead of $H_2$ in the main text because there must be a significant amount of oxygen in the metallic core considering the case of Earth[1] and oxygen solubility in the metallic core further increases with increasing temperature and pressure[2]. The oxygen can be from either $H_2O$ or silicate, or both. Recent equilibrium chemistry models have shown that during the magma ocean stage, $H_2O$ can be produced by redox-reactions between primordial hydrogen and mantle melts[3]. For example, the reactions FeO + $H_2$ <—> Fe + $H_2O$ and $SiO_2$ + $H_2$ <—> Si + $H_2O$. Although it is quite likely that H and O partition in different ratios between silicate and metal instead of a constant value of 2:1, we believe that $H_2O$ is a better starting point than pure H in terms of constraining the mass-radius relationship as the latter completely ignores the role of oxygen. Also, the effect of $H_2O$ on reducing the silicate density and solidus is much better documented than that of H[4].

## Comparison with Earth case

Recent studies have provided evidence supporting the idea that the Earth's core hosts the majority of Earth's water[5-7]. However, there remains an inherent uncertainty surrounding this conclusion due to the unknown timing of water delivery to Earth. Specifically, it is unclear whether it was delivered during main accretion or during the late veneer after the Earth's core had already formed. Additionally, the late veneer scenario itself is debated[8] and even if it took place, the compositions of late veneer materials remain uncertain[9].

For exoplanets that formed with significant water content (tens of water mass fractions), the late addition of water is not an important source. During their formation, large-scale melting allows the iron to separate from the hydrated silicates to form a core. Iron melt droplets that sink through the silicates to settle at deeper depths, equilibrate with the hydrated silicates. This equilibration allows water to efficiently partition between iron and silicates over a range of pressures. Hence, water-rich worlds naturally incorporate the majority of water within their cores, depending on planetary mass (Fig. 4). One key question is whether any water stored deep in the iron cores of planets can eventually find its way back to the mantle and, potentially, even to surface reservoirs. This aspect will be explored in the next section.

## Most of water may be locked in the core

During the evolution of a planet, the water at the surface and the mantle can be in equilibrium as long as their interface allows rocks to be molten. As water can be lost to space from the atmosphere and the mantle melt fraction decreases during the cooling, water from the mantle can replenish the surface reservoir[10]. At the core-mantle boundary, this compositional coupling is different. Replenishing the mantle reservoir with water from the core is extremely difficult for two reasons. First, metal-silicate partition coefficients of water increase with decreasing temperature (Extended Data Fig. 3). As a planet cools, more water may tend to be enriched in the core instead of replenishing the mantle reservoir. Second, element diffusion is sluggish in solid at high pressures. A thin solid layer of kilometers scale at the CMB would effectively separately the core and mantle in terms of water[11,12]. Hence, core and mantle are likely not in compositional equilibrium over the long-term evolution of a planet, as is the case for Earth.

**Limitations**

The mass-radius relations presented in our study are valid for warm and hot worlds with equilibrium temperatures above 400 K. For more temperate planets that do not host global magma oceans, core and mantle may not be in equilibrium, as described above. Hence, while the mantle reservoir may degas as it cools, the core will keep most of its initial budget of H and O over its lifetime. In addition, the surface water reservoir is subject to atmospheric loss[13,14]. Atmospheric loss and the cooling of the magma ocean both drive outgassing from the mantle reservoir. Thus, hot and close-in planets may have water-depleted mantles in equilibrium with their atmospheres; their cores, however, will not be depleted as cores and mantles will be out of equilibrium. In consequence, inferred water budgets could be even higher compared to the model D that assumes a planet in global equilibrium. Further research is needed to explore the distribution of water within temperate planets and hot planets that underwent significant atmospheric erosion (see the case of 55 Cnc in main text).

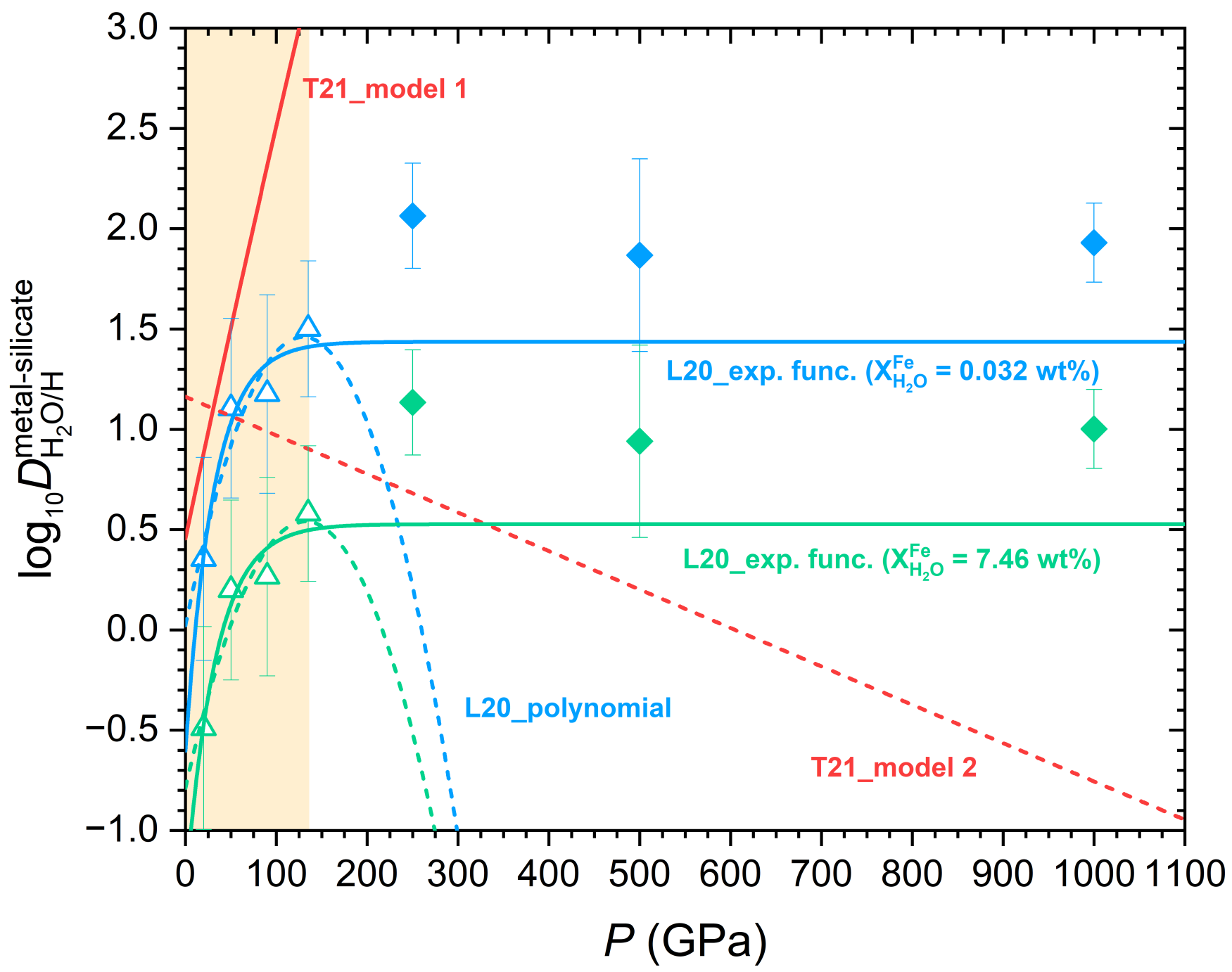

**Supplementary Fig. 1. A comparison of partition coefficients (*D*) of $H_2O$/H between metalllic and silicate melts.** Red solid and dashed lines are the extrapolated experimental values of $D_H$ based on two models describing the temperature and pressure dependence of $D_H$ reported in Tagawa et al. (2021)[7]. Open triangular symbols represent computational data ($D_{H_2O}$) from Li et al. (2020)[5] and the corresponding solid and dashed lines indicate the best fits using an exponential function and a second-degree polynomial function, respectively. Solid diamond symbols are our computational results ($D_{H_2O}$). The shaded region in light orange represents the pressure range that previous data covers.

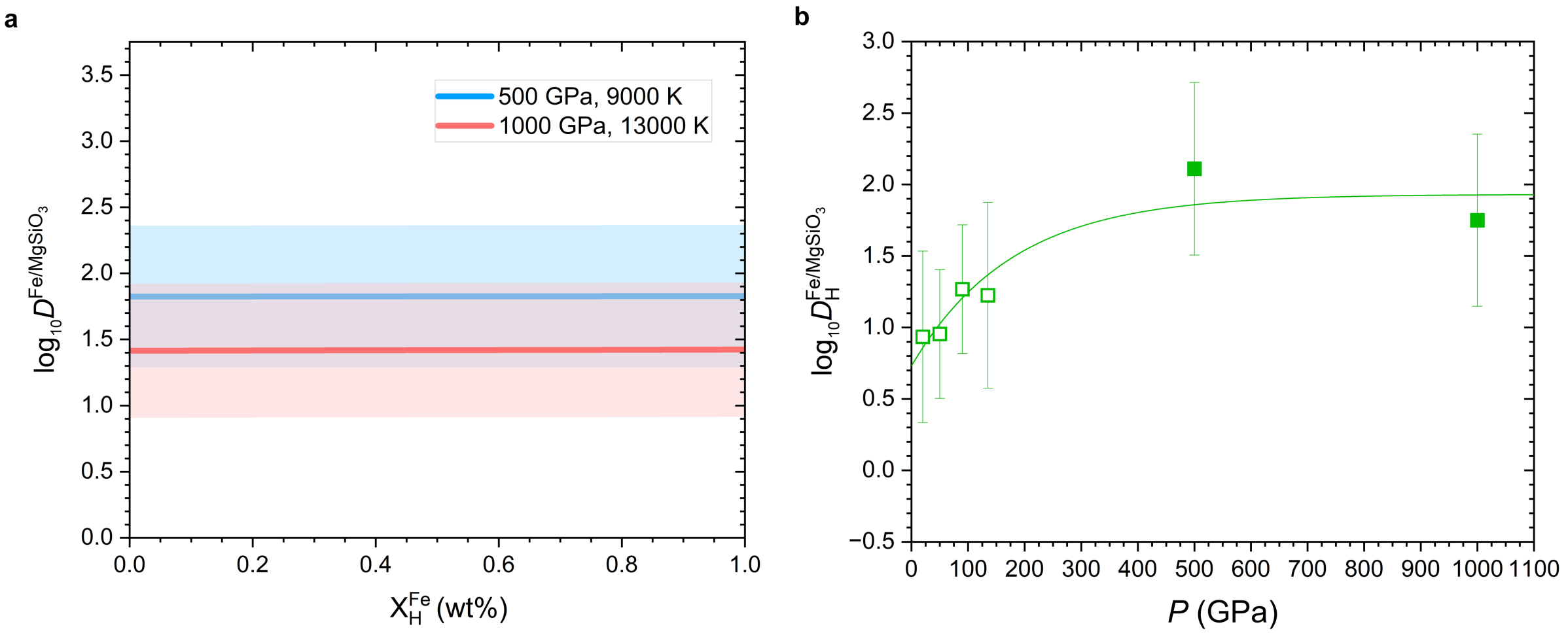

**Supplementary Fig. 2. Partition coefficients of H.** Calculated partition coefficients of H between iron and silicate melts ($D_{\mathrm{H}}^{\mathrm{Fe/MgSiO_3}}$) as a function of its concentrations in iron melt ($X_{\mathrm{H}}^{\mathrm{Fe}}$) **(a)** and as a function of pressure (*P*) **(b).** Partition coefficients are calculated as the ratio of weight fractions of hydrogen in metal and silicate. **a**, The errors of $D_{\mathrm{H}}^{\mathrm{Fe/MgSiO_3}}$ are represented by the shaded area with the same color as those of the respective lines. **b,** Open symbols are previous computational data[5]. Solid symbols represent our results. $D_{\mathrm{H}}^{\mathrm{Fe/MgSiO_3}}$ is parametrized as a function of *P* (in GPa) with the equation: $\log_{10} D_{\mathrm{H}}^{\mathrm{Fe/MgSiO_3}} = 1.93 - 844/(1 + \exp((P + 1168)/178))$.